\documentclass [notitlepage]{report}
\usepackage{epsfig}
\usepackage{amsmath}
\usepackage{maize}

\begin{document}

\title{The nonrelativistic frequency dependent electric polarizability of a bound particle}
\author{M.A. Maize and Michael Williams*\\Department of Physics\\Saint Vincent College\\300 Fraser Purchase Rd\\Latrobe, PA 15650}
\date{}
\maketitle

\begin{abstract}
In this work, we will calculate and analyze the frequency dependent electric polarizability
of a particle which is bound by a one-dimensional delta potential.  A perturbation technique
based on the work by Nozi\'{e}res will be employed to calculate our expression.  Comparison
will be made with references which have discussed work comparable to ours. \\ \\
\end{abstract}

I.  Introduction: \\ \\

The study of electric polarizability has been important in many scientific areas for a long time.  Electric polarizability is essential in understanding the electromagnetic properties of atoms, nuclei and more.  The study of the electric polarizability is strongly related to the study of the indices of refraction and dielectric constants.  In the introduction of ref\@.(1), Glover and Winhold give a nice summary of the many applications of the elctric polarizability in the area of low-energy interactions of atomic systems with each other and with electromagnetic fields$^1$.

The study of both the static and frequency dependent electric polarizabilities provide knowledge of many systems in regard to how they interact with static and frequency dependent electric fields.  In our previous work$^{2,3}$, we emplyoed the method of Dalgarno and Lewis$^4$ to calculate the static electric polarizability of a particle bound by a delta potential in both the non-relativistic and relativistic cases.  It is then logical to proceed with the frequency dependent case in both the non-relativistic and relativistic cases.  In this paper, we present the non-relativistic frequency dependent problem.  Our next step will be the study of the problem of the relativistic frequency dependent electric polarizability.

The problem of the frequency dependent electric polarizability received attention since early last century$^5$.  More recently Davydov discussed the elementary quantum theory of dispersion in the case of the polarizability of a quantum system$^6$.  C K Au derived a closed form for the dynamic multipole polarizability$^7$.  Postma derived an expression for the frequency dependent electric polarizability of the one-dimenstional hydrogen atom with delta function interaction$^8$.  The method used in refs\@.(6-8) to calculate the electric polarizability is limited to the study of the interaction between the charge of the system and an external electric field.  The technique that we use$^9$ is applicable to a wide variety of problems in addition to the electric polarizability.  One of the applications Nozi\'{e}res discusses in his book$^9$ is the study of the dielectric constant of an electron gas.  In our study we calculate and analyze the separate expressions $\alpha(\omega)$ and $\alpha(-\omega)$ ($\alpha$ is the polarizability) while such separation does not appear in refs\@.(6-8).  In refs\@.(6,7), the technique of calculating the electric polarizability and examples are presented.  However the problem of the electric polarizability in the case of a delta potential is not one of the subjects in refs\@.(6,7).  In ref\@.(8), the problem of the polarizability in the case of delta potential was studied and compared to the polarizability of the hydrogen atom.  Our work shows many differences with the results of ref\@.(8) and such differences will be discussed in section IV.  In the next section, we give a summary of Nozi\'{e}res method.  We follow this, with the presentation of our problem and the solutions which we obtained.  Next, we present our graphs and analyses.  Finally, we will have concluding remarks and appendices (A) and (B) for illustrating some needed details regarding necessary integrations and part of the differences between our results and the results of ref\@.(8).\\ \\

II.  Perturbation and Nozi\'{e}res method: \\ \\

The problem in essence is determining the response of some system due to the interaction of the system with an external perturbation varying in both space and time$^9$.  The system is originally in its ground state at $\psi_{0}$.  The hamiltonian of the ground state is H$_0$.  The system while in its ground state interacts with some external field producing an interaction potential given by \mbox{\boldmath $A$}F(t).  \mbox{\boldmath $A$} is a hamiltonian operator that operates on the wave function of the system$^9$.  Its time dependence is absorbed completely in the real scalar factor F(t).  The ``response'' of the system is nothing more than the average value of a certain operator, \mbox{\boldmath $B$}$^9$, which can be the current, the magnetic moment$^9$, the electric dipole moment, etc.  The average is taken relative to the wave function $\psi(t)$.  $\psi(t)$ is given by: \begin{equation} \psi(t) = e^{iH_{0}t}\psi_s(t) \ \ \ \ , \end{equation} where $\psi_{s}$(t) is the solution of the following equation: \begin{equation} i\frac{\delta\psi_{s}(t)}{\delta t} = (H_{0} + \mbox{\boldmath $A$}F(t))\psi_{s}(t) \ \ \ \ \ \ . \end{equation} The function $\psi$ is then calculated to the first order in \mbox{\boldmath $A$} via the expansion$^9$ \begin{equation} \psi = \psi_0 + \psi_1 + \ldots \ \ \ \ \ \ , \end{equation} where $\psi_1$ satisfies the following equation: \begin{equation} i\frac{\delta\psi_1}{\delta t} = \mbox{\boldmath $A$}(t)F(t)\psi_0 \ \ \ \ \ \ \ , \end{equation} where \mbox{\boldmath $A$}(t) is given by: \begin{equation} \mbox{\boldmath $A$}(t) = e^{iH_0t}\mbox{\boldmath $A$}e^{-iH_0t} \ \ \ \ \ \ . \end{equation}  Only the components $\psi_0$ and $\psi_1$ are used to find the average value of \mbox{\boldmath $B$}.  In the equations cited from ref\@.(9), $\hbar$ is set equal to 1.  The response of the system is then measured by the function $\chi_{BA}(\omega)$ which is then given by$^9$: \begin{equation} <\mbox{\boldmath $B$}> - \mbox{\boldmath $B$}_0 = \chi_{BA}(\omega) exp[(-iw + \mu)t] \ \ \ \ \ \ , \end{equation} where \begin{equation} \mbox{\boldmath $B$}_0 = <\psi_0|\mbox{\boldmath $A$}|\psi_0> \ \ \ \ \ , \end{equation} and $<$\mbox{\boldmath $B$}$>$ is the average of \mbox{\boldmath $B$}.  $\omega$ is the frequency and the factor $e^{\mu t}$ is introduced to set some restrictions on the behavior of F(t)$^8$.  After some manipulations, the final expression of $\chi_{BA}(\omega)$ (the response function) is given as$^8$: \begin{equation} \chi_{BA}(\omega) = \sum_{n} [\frac{<\psi_0|\mbox{\boldmath $B$}|\psi_n><\psi_n|\mbox{\boldmath $A$}|\psi_0>}{\omega - \omega_{n_0} + i\mu} - \frac{<\psi_0|\mbox{\boldmath $A$}|\psi_n><\psi_n|\mbox{\boldmath $B$}|\psi_0>}{\omega + \omega_{n_0} + i\mu}] \ \ \ \ , \end{equation} where  $\omega_n - \omega_0 = \omega_{n_0}$.  The functions $\psi_n$'s represent a complete set of eigenfunctions of the hamiltonian $H_0$, of energy $\omega_0$.\\ \\

III.  Model and The Polarizability: \\ \\

The hamiltonian $H_0$ which is associated with the ground $\psi_0$ and the ground state energy $E_0$ is given by: \begin{equation} H_0 = -\frac{\hbar^2}{2m} \frac{\delta^2}{\delta x^2} - g\delta(x) \ \ \ \ \ \ , \end{equation} where g is the strength of the potential and m is the mass of the particle occupying the state $\psi_0$.  The solution of the Schr\"{o}dinger equation $H_0\psi = E\psi$ will produce a bound state which is given by: \begin{equation} \psi_0(x) = \sqrt{k_0}e^{-k_0|x|} \ \ \ \ \ \ , \end{equation} and a ground state energy $E_0$ which is given by: \begin{equation} E_0 = -\frac{\hbar^2k^2_0}{2m} \ \ \ \ \ \ , \end{equation} where $k_0 = \frac{mg}{\hbar^2}$.  In addition to the ground state $\psi_0$, the Schr\"{o}dinger equation will produce free particle states.  We will not give here the expression of the free particle states since they will not be used in our calculation.  The expressions can be found in many text books.

In calculating the electric polarizability, the excitations are due to the electric dipole.  The interaction between the external electric field and the charge of the system is then equal to -q$\varepsilon$x, where q is the charge and $\varepsilon$ is the strength of the field.  To calculate the electric polarizability with the use of eq\@.(8), we need to calculate the matrix elements $<\psi_0|xG(\pm\omega)x|\psi_0>$ where $G(\pm\omega)$ are given by: \begin{equation} G(\pm\omega) = G_0(\pm\omega) + G_0(\pm\omega)V(x)G(\pm\omega) \ \ \ \ \ , \end{equation} with V(x) being the attractive delta potential.  $G(\pm\omega)$ and $G_0(\pm\omega)$ are given in terms of T, $H_0$, $\omega$ and $E_0$ by the following expressions: \begin{equation} G(\pm\omega) = \frac{1}{E_0 \pm \omega - H_0} \ \ \ \ \ \ \ , \end{equation} and \begin{equation} G_0(\pm\omega) = \frac{1}{E_0 \pm \omega - \mbox{\boldmath $T$}} \ \ \ \ \ \ \ \ , \end{equation} where $\mbox{\boldmath $T$} = -\frac{\hbar^2}{2m} \frac{\delta^2}{\delta x^2}$.  To calculate the matrix element $<\psi_0|xG(\pm\omega)x|\psi_0>$, we use the completeness of the free particle states $|k>$ and $|k'>$ $[|k> = \frac{1}{\sqrt{2\pi}}e^{ikx}]$ to find the matrix elements $<\psi_0|x|k'>$, $<k'|G(\pm\omega)|k>$ and $<k|x|\psi_0>$, then do the necessary integrations over k and k'. The derivations of the matrix elements of $G(\pm\omega)$ in terms of the matrix elements of $G_0(\pm\omega)$, $\omega$ and the necessary constants is straightforward and we will give the necessary integrations in appendix A.  It is also important to stress that no approximation was made.

The two terms on the r\@.h\@.s\@. of eq\@.(8) will contribute equally to the static electric polarizability.  The complete behavior of $\alpha(\omega)$ and $\alpha(-\omega)$ for $\omega = 0$ and $\omega \neq 0$ will be obtained from the expressions we derived with the aid of eq\@.(8).  $\alpha(\omega)$ corresponds to the first term on the r\@.h\@.s\@. of eq\@.(8) while $\alpha(-\omega)$ corresponds to the second term on the r\@.h\.s\@. of eq\@.(8).  We set in our equations $B = -E_0$, and we take $\hbar$, $q$ and $m$ to be equal to one.  $\alpha(\omega)$, which we obtained is given by:
\begin{equation}
\alpha(\omega < B) = \frac{1}{k^{2}_0(k_0 + k)^2} + \frac{2}{k_0(k_0 + k)^3} + \frac{2}{(k_0 + k)^4} \ \ \ \ \ \ \ \   , 
\end{equation}
and
\begin{equation}
\begin{split}
\alpha(\omega > B) =& -\frac{1}{k^2_0(k^2_0 + \Lambda^2)} - \frac{2}{(k^2_0 +
\Lambda^2)^2} - \frac{8k^2_0}{(k^2+\Lambda^2)^3} \\
&+ \frac{16k^4_0}{(k^2_0 + \Lambda^2)^4} + \frac{i 16 \Lambda k^3_0}{(k^2_0 + \Lambda^2)^4} \ \ \ \ , 
\end{split}
\end{equation} where $k^2 = 2(B - \omega)$ and $\Lambda^2 = 2(-B + \omega)$. 
The expression for $\alpha(-\omega)$ is given by:
\begin{equation}
\alpha(-\omega) = \frac{1}{k^2_0(k_0 + k')^2} + \frac{2}{k_0(k_0 + k')^3} 
+ \frac{2}{(k_0 + k')^4} \ \ \ \ \ \ \ \ , 
\end{equation} 
where ${k'}^2 = 2(B + \omega)$. \\ \\

IV.  Graphs and analyses: \\ \\

Studying the frequency dependent electric polarizability in terms of
$\alpha(\omega)$ and $\alpha(-\omega)$ does indeed help in better understanding
the response of the system for the various values of $\omega$ i\@.e\@. $\omega < B$
and $\omega > B$.  In the work of ref\@.(8), $g$ was set equal to one, with m and $\hbar$ set equal to one leading to $E_0 = - \frac{1}{2}$, the frequency dependent electric polarizability
$\alpha(\omega)^8$ was then written as a real part $\alpha_r(\omega)^8$ and an imaginary
part $\alpha_i(\omega)^8$.  $\alpha_r(\omega)$ was then written for $\omega < \frac{1}{2}$
and $\omega > \frac{1}{2}^8$.  $\alpha_r(\omega)$ for $\omega < \frac{1}{2}^8$ corresponds
to $\alpha(\omega < B) + \alpha(-\omega)$ for $\omega < B$ in our work.  $\alpha_r(\omega)$
for $\omega < \frac{1}{2}$ is given by$^8$:
\begin{equation}
\alpha_r(\omega) = \frac{2 - \omega^2 - \sqrt{1 + 2\omega} -
\sqrt{1 - 2\omega}}{\omega^4} \ \ \ \ \ \ . \ \
\end{equation}
To make direct comparison with our work; we set $B = \frac{1}{2}$ in both eqs\@.(15) and (17).
After the substitution, we get the following for $\omega < \frac{1}{2}$:
\begin{equation}
\alpha(\omega < \frac{1}{2}) + \alpha(-\omega) = \frac{1}{(1 + \sqrt{1-2\omega})^2}
+ \frac{2}{(1 + \sqrt{1 - 2\omega})^3} + \frac{2}{(1 + \sqrt{1 - 2\omega})^4}
+ \omega \rightarrow -\omega \ \ \ \ \ \ , 
\end{equation} 
Where $\omega \rightarrow -\omega$ is the first three terms on the r\@.h\@.s\@. of eq\@.(19) with $-\omega$ replacing $\omega$.  The static electric polarizability is obtained in our expression by simply setting  $\omega = 0$ in eq\@.(19).  To get the static electric polarizability in ref\@.(8) an expansion in the case of $\omega < < 1$ has to be performed on the r\@.h\@.s\@. of eq\@.(18).  Such expansion in ref\@.(8) produces $\alpha_r(\omega) = 1.25(1 + 2.1\omega^2 + \ldots)$.  In our work an expansion in the case of $\omega < < 1$ on the r\@.h\@.s\@. side of eq\@.(19) produces $1.25(1 + 1.6\omega^2 + \ldots)$.  Our expression of eq\@.(19) produces the same static electric polarizability and the same behavior at $\omega = \frac{1}{2}$ as the expression of eq\@.(18)$^8$, however the details of the two expressions are different.

Fig\@.I shows our results of the electric polarizability for $x < 1$.  $\alpha(\omega < B)$
and $\alpha(-\omega)$ contribute equally to the static electric polarizability. 
$\alpha(\omega < B)$ will keep increasing while $\alpha(-\omega)$ will decrease for larger x, with $\alpha(\omega < B)$ strongly dominating $\alpha(-\omega)$ in the vicinity of and at
$x = 1$ [fig\@.I] where $x = \frac{\omega}{B}$.  $x = 1 (\omega = B)$ represents the threshold of what we know as the photelectric effect.  As we study in our undergraduate courses, the threshold of the photelectric effect happens when the energy of the photon is exactly equal to the work function.  $B$, the value of the binding energy represents the work function. 
We also know from our study of the photoelectric effect that for $\omega \geq B$, there
is a total absorption of the photon of energy $\omega$.  Such absorption leads to the
existence of the imaginary part of $\alpha(\omega > B)$ [fig\@. III].  Since
$\alpha(-\omega)$ does not contain a term (terms) which corresponds to the absorption
of a photon by the bound state particle, $\alpha(-\omega)$ does not contain an imaginary part.  Referring to fig\@.II, we can see that for $\omega \geq 6B$, there is a resonable cancellation between $Re \alpha(\omega > B)$ and $\alpha(-\omega)$.  Such a cancellation can not be seen in ref\@.(8) since the reference does not separate $\alpha(\omega)$ and $\alpha(-\omega)$.  Our expression of the polarizability for the region $\omega > B$ is different from corresponding expression in ref\@.(8) [appendix B].  There is an agreement between our work and ref\@.(8) in the imaginary part of the polarizability.  In the high $\omega$ range (say $\omega \geq 5B$), the details of the dispersion relation disappear.  This points to the limitations of the non-relativistic problem and the necessity of studying the relativistic problem.  We like to add at the end of this section that a good discussion about the relation between the frequency dependent electric polarizability and the frequency dependent index of refraction can be found in ref\@.(6). \\ \\

V.  Conclusion: \\ \\

In this paper, we calculated the nonrelativistic frequency dependent electric polarizability
of a particle bound by a one-dimenstional delta potential with the use of Nozi\'{e}res
method$^9$.  We believe that calculating both $\alpha(\omega)$ and $\alpha(-\omega)$
helped in better illustration of the response of the system.  Our results have their differences with ref\@.(8).  Such differences were discussed in the previous section and appendix B.  At the same time, the specific problem of the polarizability in the case of the delta potential was not discussed in refs\@.(6) and (7).

We believe that the work presented in this paper will server more than one purpose.  With the work we did in refs\@.(2) and (3) and our plan to study the relativistic frequency dependent elctric polarizability in the immediate future, the work of this paper represents an essential part of studying the porblem of the electric polarizability of a particle bound by a delta potential for the static case and the frequency dependent cases while considering the non-relativistic and the relativistic problems.  We believe that our presentation is educational for both undergraduate and graduate students since it deals with important fundamentals of physics and introduces the application of useful techniques.  Finally, the results of the non-relativistic problem demonstrated in a clear way, the importance of doing the relativistc problem. \\ \\

\appendix
\chapter{}

In this appendix, we will give the method used in performed the integrations which appear when dealing with the calculation of $\alpha(\pm\omega)$.  Also we will give the result of another integration.  There were no approximations done in our work or any integration done analytically or with the use of integration tables as done in ref\@.(8).

An integration ($I_1$), which appeared in our calculation can be written as: \begin{equation} I_1 = \int_{-\infty}^{\infty} \frac{dx x^2}{(x^2 + x_0^2)^4(x^2 + a^2)} \ \ \ \ \ \ . \end{equation}  To solve for $I_1$ numberically it can be written as: \begin{equation} I_1 = (-\frac{1}{6x_0} \frac{\partial}{\partial x_0})(-\frac{1}{4x_0}\frac{\partial}{\partial x_0})(-\frac{1}{2x_0}\frac{\partial}{\partial x_0}) \int_{-\infty}^{\infty} \frac{x^2 dx}{(x^2 + x_0^2)(x^2 + a^2)} \ \ \ \ \ . \end{equation}  An integration ($I_2$) which we encounter is given by: \begin{equation} I_2 = \int_{-\infty}^{\infty} \frac{dk'}{(2\pi)(\omega + \omega_0 - E')} \ \ \ \ \ \ \ . \end{equation}  The value of $I_2$ is $-\frac{i}{\sqrt{2\Omega}}$ where $\Omega = \omega + \omega_0$ and $E^{'2} = \frac{k^{'2}}{2}$ with $\hbar$ and $m$ set equal to one.  $\omega$ and $\omega_0$ have the same definitions which we used throughout the paper. \\ \\

\begin{figure}[htbp]
\begin{center}
\input{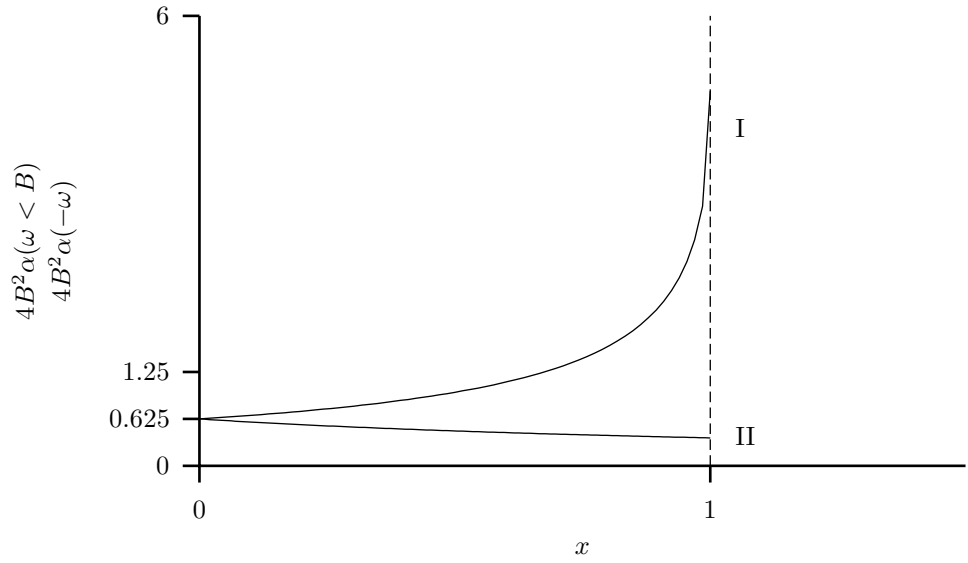}
\end{center}
\caption{Plot of $4B^2\alpha(\omega < B)$ which appears as I and $4B^2\alpha(-\omega)$ which appears as II versus $x$ where $x = \frac{\omega}{B}$, for $x(0 \rightarrow 1)$}
\end{figure}

\begin{figure}[htbp]
\begin{center}
\input{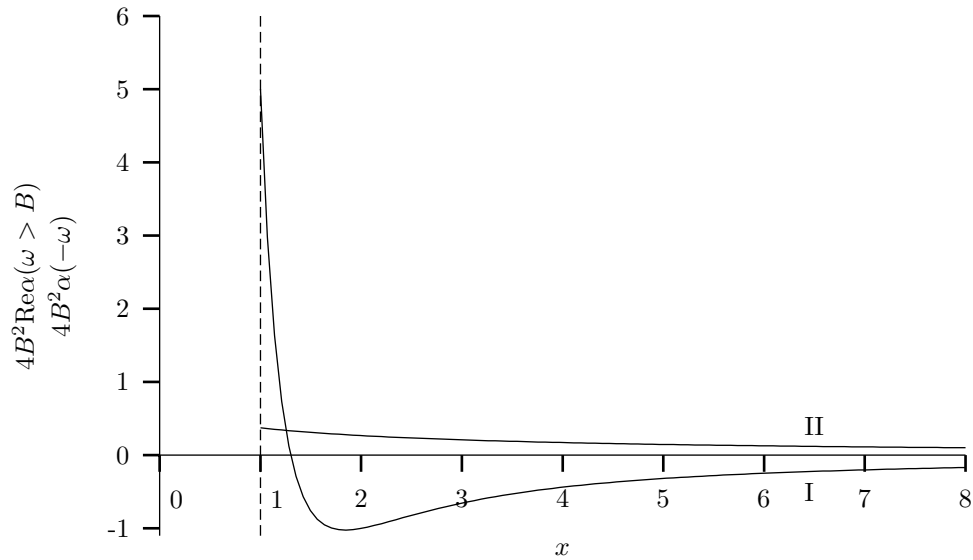}
\end{center}
\caption{Plot of the real part of $\alpha(\omega > B)$ multiplied by $4B^2$
($4B^2$Re$\alpha(\omega > B)$ which appears as I and $4B^2\alpha(-\omega)$ which
appears as II vesus $x$, where $x = \frac{\omega}{B}$, for $x > 1$}\end{figure}

\begin{figure}[htbp]
\begin{center}
\input{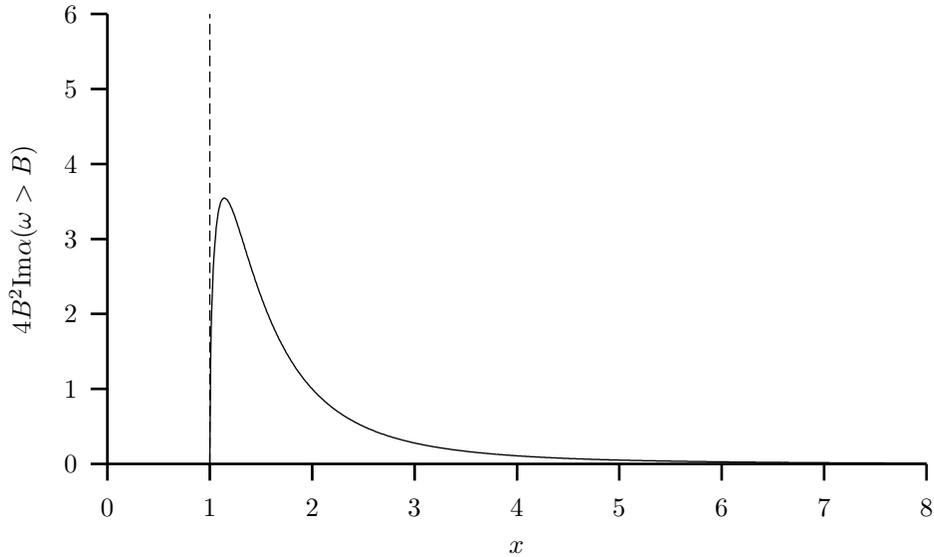}
\end{center}
\caption{Plot of the imaginary part of $\alpha(\omega > B)$ multiplied by $4B^2 (4B^2$Im$\alpha(\omega > B))$ versus $x$, where $x = \frac{\omega}{B} for x > 1$}
\end{figure}

\chapter{}

In ref\@.(8), B is equal to $\frac{1}{2}$.  Due to this, the real part of the polarizability $\alpha_r(\omega)$ is given by$^8$: \begin{equation} \alpha_r(\omega) = \frac{2 - \omega^2 - \sqrt{2\omega + 1}}{\omega^4} \end{equation}  For $B = \frac{1}{2}$, our term which corresponds to $\alpha_r(\omega)$ is given by:
\begin{equation}
\mathrm{Re}\,\alpha(\omega > B) + \alpha(-\omega) = \frac{w - 2\omega - \omega^2 - \omega^3}{2\omega^4} + \frac{1}{(1 + \sqrt{1 + 2\omega})^2} + \frac{2}{(1 + \sqrt{1 +2\omega})^3} + \frac{2}{(1 + \sqrt{1 + 2\omega})^4} \ \ \ \ . 
\end{equation}

\bigskip
\noindent
* Current address: Dept. of Physics, Carnegie Mellon University, Pittsburgh, PA 15213

\end{document}